# Growing Into an Adaptive and Reconfigurable Paradigm for Spectrum Sharing

Charles Baylis[1], Douglas Sicker[2], Austin Egbert[1], Andrew Clegg[1], Tom Brooks[1], Casey Latham[3], and Robert J. Marks II[1]

[1]Department of Electrical and Computer Engineering, Baylor University
[2]Department of Computer Science, Baylor University
[3]Keysight Technologies
Charles_Baylis@baylor.edu

*Abstract*—A significant movement from rigid use of the wireless spectrum toward adaptive and reconfigurable spectrum use has been prompted by increasing spectral crowding. Some bands have moved to an adaptive sharing model, and proposals are growing for this approach to be applied to additional bands. The process of moving from a fixed, rigid spectrum paradigm to adaptive and reconfigurable use involves maturation of policy and technology at multiple levels within the system of systems. Using the concept of Bloom's Taxonomy from the education discipline, this paper examines the development of a policy and technology progression toward a mature, adaptive and reconfigurable paradigm.

## I. Introduction

A trajectory from rigid, inflexible spectrum use toward adaptive and reconfigurable spectrum usage is now in full force. One of the earlier instantiations of large-scale dynamic sharing structures was the former television "White Space" band in the early 2010s [1]. In the mid 2010s, defense systems began to be involved in sharing, as the Citizens Broadband Radio Service (CBRS) was launched and provided the opportunity to share with Navy radars from 3.55-3.7 GHz in the United States [2]. Recently, sharing of spectrum designated for point-to-point microwave links with unlicensed devices in the 6 GHz band has been enacted. In 2024, several Automated Frequency Coordination (AFC) systems, cloud-based sharing systems that coordinate between the incumbent (point-to-point microwave links) and the new users, were approved for operation [3].

Sharing is now being considered for the 3.1-3.45 GHz band in the United States, as mandated by the Infrastructure Investment and Jobs Act of 2021 [5]. Incumbent military radar systems, and other systems, may be required to share with wireless communication systems. The National Spectrum Strategy Implementation Plan prescribes study of this band, with a report due in 2026 [6].

The move toward dynamic allocation of spectrum is clear; however, this major change in spectrum use lacks a roadmap, which has hampered its adoption. Without a well-reasoned roadmap, there is too much complexity at too many levels to ensure the timely and appropriate evolution towards a meaningful and desirable future spectral designation. In building a new spectrum paradigm, it is necessary to consider the evolution of the paradigm; that is, the process through which the paradigm can move from present approaches to fulfill desired future methods of spectrum sharing. We discuss a method for growing an adaptive and reconfigurable spectrum-use paradigm by applying well-known educational methods (Bloom's Taxonomy) to technology and policy development.

## II. Multi-Level Adaptivity and Reconfigurability

As a prerequisite to determining next steps, the desired future spectrum-use paradigm must be defined. To allow real-time spectrum sharing, adaptivity and reconfigurability are needed at all levels. At the highest level, informed policy decisions must be made in real-time to permit spectrum sharing. At the lowest level, reconfigurable circuits allow intelligent optimization of performance upon switching operating frequency. The concept of adaptive and reconfigurable spectrum policy can be envisioned similar to driving an automobile on roadways, as shown in Fig. 1.

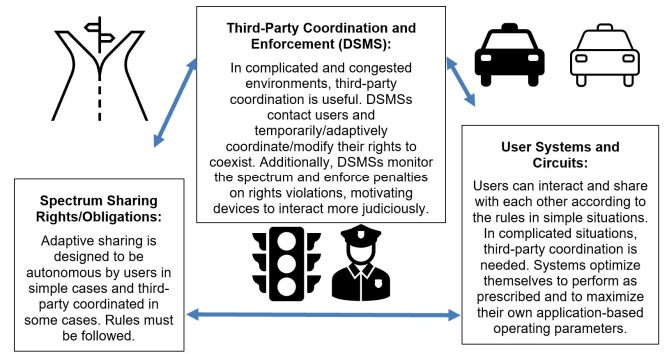

Fig. 1. Description of a possible adaptive and reconfigurable spectrum usage paradigm

In roadway sharing, drivers and vehicles follow protocols and policies for road use that avoid crashes and allow each car to maximize its objectives, while maintaining safety. In spectrum sharing, similar protocols and policies are needed to enable mutually beneficial, real-time spectrum sharing. Just as traffic signals are needed in some conditions, but drivers can often autonomously change lanes in less congested scenarios, a third-party spectrum coordinator may be useful in many situations, but is not necessary in all scenarios. A Dynamic Spectrum Management System (DSMS), paralleling law enforcement officers and traffic signals, must be developed for real-time coordination between devices in more congested scenarios where collisions are likely if autonomous operation remains. In less congested scenarios, autonomous, unsupervised interaction between systems may be most useful. Finally, devices, circuits, and systems are required that can reconfigure their systems to automatically optimize performance upon changing operation to different spectrum.

To build this new paradigm, we propose the application of educational methods to policy and technology maturation.

### III. Applying Bloom's Taxonomy to Growing a Spectrum Sharing Paradigm

First published in 1956, Bloom's Taxonomy is a common guide for educators on the development of cognitive capabilities [7]. Bloom lists a progression of six stages that demonstrate a progression from very simple learning toward full cognitive capability: (1) Knowledge, (2) Comprehension, (3) Application, (4) Analysis, (5) Synthesis, and (6) Evaluation [7]. This progression of stages during the educational process is considered by evaluating the progression of students through the learning and maturation process. Fig. 2 shows a visualization of Bloom's Taxonomy.

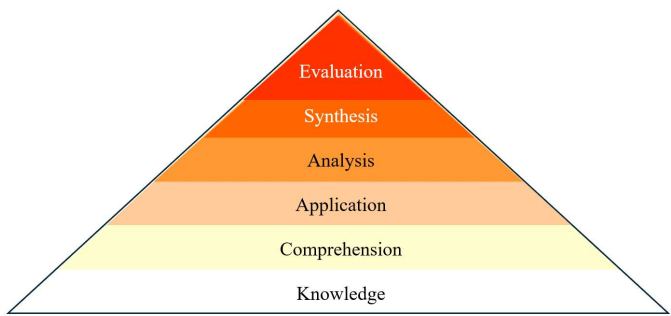

Fig. 2. Bloom's Taxonomy [7], often pictured in the literature as a triangle with cognitive development progressing from bottom to top

Given the analogy of Fig. 1, which compares the future paradigm of adaptive and reconfigurable spectrum sharing to vehicles sharing a roadway, the problem under consideration is the maturation of spectrum use toward autonomous sharing. This maturation processes through the different levels of the Fig. 2 Bloom's Taxonomy structure, from bottom to top. Based on the road analogy, this is similar to training a driver. A driver is not trained instantaneously; rather, it is the primary and secondary education process, along with specialized, experiential driver education, that fully equips the driver to be successful on the road. First, elementary-school children are taught the "Knowledge" level: they memorize math facts, numbers, and letters, but know little about how to apply them. In "Comprehension", students learn how to understand what they are reading or to form conclusions based on experiences. In "Application", students learn how to solve problems that require design using basic principles. In "Synthesis", students are required to weave concepts together across multiple disciplines to create new material (This paper is actually an example of "Synthesis" because it brings together a technological concept with an educational concept). A driver, through "Synthesis", may read traffic reports and road maps, and choose a fastest path to the desired destination. Finally, "Evaluation" requires informed judgment upon presented material, results or experiments. At the "Evaluation" level, a person completely skilled at driving can assess the evolving (and potentially dangerous) surrounding environment, along with road rules and destination information, and quickly make a real-time decision about next actions that can promote efficiency, provide safety, and even save lives.

During the evolution through this Taxonomy, reasonable expectations must be placed on spectrum use. Just as a child is expected to make mistakes and mature as a result, it should be noted that there will be some issues encountered along the way. As system designers provide improvements in response to these challenges, cognitive capability will mature and become more robust. Machine Learning can also be used to enhance performance as a result of lessons learned. During the evolution of the taxonomy, filters, guardbands (just as seatbelts and bumpers), and auctions will still be useful in protecting spectrum usage.

A similar process must take place for adaptive and reconfigurable spectrum sharing. A high-level example of spectrum policy and technology development using Bloom's Taxonomy is provided as follows:

1) Knowledge: Spectrum is pre-assigned and is rigid. Individual wireless devices operate strictly in pre-assigned bands without any evaluation of performance or consideration of other opportunities.
2) Comprehension: A cause-and-effect relationship can be derived between spectral assignment options and system performance.
3) Application: A wireless device can move to a different spectrum band in the presence of interference, applying the understanding of interference to avoid it.
4) Analysis: The usage of the spectrum by a system, or by a system of systems, can be evaluated, and an assessment of potentially improved allocation and sharing options can be performed.
5) Synthesis: Highly "creative" (or advanced artificial intelligence and machine learning) techniques can be marshalled to investigate which band combinations will be optimal for given groups of wireless devices.
6) Evaluation: Wireless devices can, in real-time, evaluate their own performance (and the performance of a system of systems) in a given sharing scenario or device setting, and can then re-optimize the device scenario or setting to improve performance. This is a "closed loop" optimization of device and system performance.

Given these different levels of Bloom's Taxonomy, it is possible to develop specific roadmaps for spectrum policy and for different technology layers to process through the Taxonomy.

### IV. Developing a Policy Roadmap

The policy roadmap should be created for policies to allow spectrum sharing at the different maturity levels indicated by Bloom's Taxonomy. A candidate policy roadmap might appear as follows:

1) Knowledge: Spectrum is pre-assigned and is enforced without any level of practical flexibility or adjustment.

2) Comprehension: Detection and analysis of interference events (or potential interference events) is possible. This level of Bloom's Taxonomy can be reached by evaluating presently used policy structures and examining benefits and drawbacks of the different structures.
3) Application: In the presence of potential interference events, policy structures are created that can allow correction of an interference scenario by relocating one of the systems to a different band or by creating an interference mitigation scenario, working within a policy framework. Based upon the "Comprehension" studies, these potential band adjustments to ease interference allow certain case-based modifications of the rigid policy structure to reduce interference occurrences. While not fully adaptive and reconfigurable, this stage of Bloom's Taxonomy is the first stage where any adaptivity elements are included in spectrum policy.
4) Analysis: Policy structures make allowance for evaluating the performance of a presence of interference and adjusting policy to compensate for reduced performance due to interference. To reach this stage of Bloom's Taxonomy, closed-loop evaluation of performance after adaptation is necessary. The development of the ability to use spectrum in a closed-loop fashion: assessing the impact of different policy structures and adjusting to optimize the impact on involved users, allows this level to be reached.
5) Synthesis: Spectrum policy structures incorporate intelligent techniques, such as artificial intelligence and machine learning, to maximize the performance of a system of systems based on changing requirements and objectives. In this level of Bloom's Taxonomy, spectrum sharing becomes less case-by-case basis and more automated, relying on intelligent methods to sense the need for relocation by different systems and then assessing the results toward future potential relocations.
6) Evaluation: Policy adapts in real-time to pre-assign, then re-assign, bands based on closed-loop evaluation of device performance metrics and coexistence needs, either with or without the assistance of a DSMS. To reach this final level of Bloom's Taxonomy, spectrum use should be fully adaptive (fully able to respond to other users, DSMSs, and assessments of a user's own performance), and reconfigurable (capable of adjusting in reasonable, prescribed ways to meet system-of-system needs), with full evaluation and response capabilities. If a first reconfiguration does not optimize performance, the reconfiguration strategy should be changed "on the fly" to ensure optimal performance at all times.

Fig. 3 shows the Bloom's Taxonomy for policy, summarizing the roadmap for needed policy development. Given that the present scenario involves the "Knowledge" part of the taxonomy, with slight infusions of "Comprehension" and "Application", this discussion has described how the roadmap should be developed to move state-of-the-art policy all the way through the Taxonomy.
In developing the detailed roadmap for policy development, the concepts of rights and integrating economic considerations into real-time spectrum sharing must be investigated. Using a DSMS only when necessary and developing the ability to provide useful peer-to-peer interaction for sharing between individual wireless devices should be prioritized. However, rules for interaction should be developed only at the level necessary to preserve fairness of interaction and ensure stimulation for successful sharing. Additionally, regulation must preserve public-service spectrum usage (such as government radar or public-safety applications) and assign it appropriate priorities along with economically driven usage (such as commercial wireless).

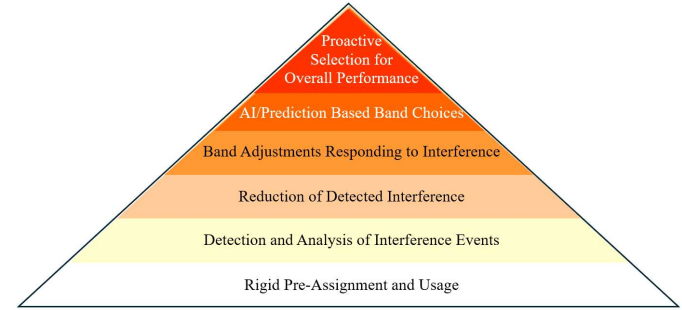

Fig. 3. Bloom's Taxonomy for wireless spectrum policy

## V. Developing Technology Roadmaps for Wireless System Layers

The technology roadmaps for each layer of a wireless system through Bloom's Taxonomy must be constructed. The following is a sample approach of a technology roadmap, as shown in Fig. 4:

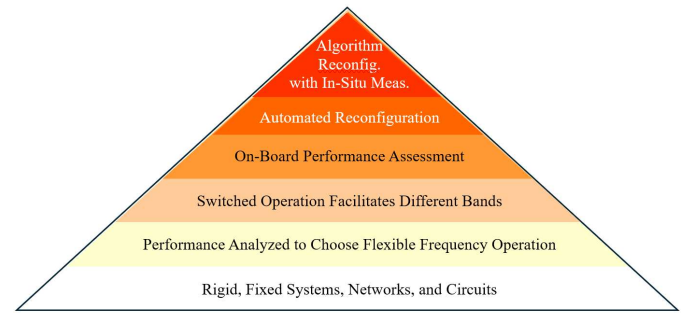

Fig. 4. Bloom's Taxonomy for wireless spectrum technology development

1) Knowledge: Hardware and networks are deployed without the ability to reconfigure upon changing in frequency. The hardware and networks are fixed. While "knowledge" of how the device performs is available, no change or "learning" is available.
2) Comprehension: Technical analysis of components can be performed to choose, with a fixed device, frequency options that are most amenable to maximizing device performance. For example, a power amplifier can be operated within a list of frequency ranges where the output power is maximized, allowing greatest radar range. While the device itself cannot adapt, it can "comprehend" its performance to make decisions.
3) Application: The ability to make choices between different technology options within a given system is available. While the device, circuit, and/or network are not physically reconfigurable at this stage of the taxonomy, the ability to make selections allows the first level of "adaptivity" and "reconfigurability" to be evident in hardware. In hardware implementation, this might involve switching between different fixed options.
4) Analysis: On-board capability to measure performance is included in the circuit or system. For example, the ability to perform in-situ measurements within an array is the first step in considering re-calibration of the array for changing of beam patterns, frequencies, or mutual coupling. While the on-board capability is not yet available to actually be used in a reconfiguration of the ultimate level, the closed-loop capability to assess system performance will be a key enabler in reaching the "Evaluation" level. The reconfiguration of devices becomes less case-by-case and is continuously reconfigurable through an automated process. At this stage, AI and ML can be used to inform these reconfigurations for maximization of circuit or system performance.
6) Evaluation: In-situ measurement capability is integrated with continuous, algorithm-based circuit/system/network reconfiguration to allow improvement of performance through the algorithm. This hardware will support the ability to flexibly use the spectrum while maximizing performance. The circuit or system will learn from its performance in future optimizations.

More specific technology roadmaps will be needed at different layers of the wireless system "stack". For example, a roadmap will be needed at the network level as well as at the circuit level. At the network level, it is desired that networks change from rigid, pre-assigned networks to flexible networks that can be formed between different devices to ensure optimal coexistence. At the circuit level, the roadmap must traverse a trajectory from fixed circuitry to reconfigurable circuitry, reconfigurable by AI and ML assisted algorithms, optimizing performance based on accurate, on-board "in-situ" microwave measurements.

The first step in plotting the way forward is to identify the "big picture", that is, the needed overall end goal, and then to identify the current taxonomy state of each layer in progressing to that goal. This will provide concrete direction for planning and designing future research initiatives at all layers of the stack. Example layers of the technology stack include the system-of-systems, system, network, circuit, and device levels. Additionally, measurement techniques and AI/ML approaches must also be developed through Bloom's Taxonomy.

## VI. Next Steps

Based on the roadmaps defined for policy and the different technology layers, research directions should be organized based on (a) immediate impact toward the upper taxonomy levels, and (b) successful long-term accomplishment of reaching the top of Bloom's Taxonomy. When the top of Bloom's Taxonomy is reached, a successful paradigm shift will have occurred. The goal is to get as far up Bloom's Taxonomy in each area as quickly as possible. Efforts should be parallelized between policy and different layers of the technology stack that will work together: policy and technology paths to climb the taxonomy should be traversed in parallel and should inform one another.

The mapping of paths for policy and technology will achieve the following results: (1) creating a top-level motivated, national approach to wireless spectrum research in policy and technology, and (2) mapping of the work to different sectors that influence spectrum usage. The result of using Bloom's Taxonomy to systematically create a research sequence toward adaptive and reconfigurable technology will be a multidisciplinary co-research mapping to determine the intertwining of research efforts and initiatives. Many research initiatives tend to be "bottom-up". While not all bad, one tends to view research problems from their own "foxhole" rather than thinking at a macroscopic level. The idea of building policy and technology roadmaps through Bloom's Taxonomy, based on the desired end result, will identify multidisciplinary research paths and objectives to accomplish macroscopic goals. This will orchestrate how spectrum engineers and policy makers must work together to achieve a fully adaptive and reconfigurable spectrum-use paradigm. With a clearly defined strategy that will unfold over time, a "big picture" set of goals will prevail and permeate the solution of "small picture" problems to be solved by disciplinary specialists.

## VII. Conclusions

A method for determining how to provide a paradigm shift in wireless spectrum usage from fixed, inflexible allocations and systems to adaptive, reconfigurable allocations and systems has been described. This method uses Bloom's Taxonomy, a trajectory for development of human cognition, to discern how policy and technology must advance through different stages to achieve fully adaptive and reconfigurable performance. While a comprehensive taxonomy or roadmap is necessary, it is not sufficient, and other factors must come into play. Coordinated research, funding, workforce development, and more must be aligned with the chosen Taxonomy-based roadmap to provide ultimate success. The Taxonomy provides structure and guidance for organized research and development for the needed adaptive and reconfigurable paradigm.